\documentclass[utf8]{FrontiersinHarvard} 

\usepackage{url,hyperref,lineno,microtype,subcaption}
\usepackage[onehalfspacing]{setspace}
\usepackage[normalem]{ulem}

\usepackage{color}

\usepackage{booktabs}

\def\keyFont{\fontsize{8}{11}\helveticabold }
\def\firstAuthorLast{Morrell {et~al.}} 
\def\Authors{Corinne Morrell\,$^{1,2,*}$, Mark P. Rast\,$^{1,2}$, Shah Mohammad Bahauddin\,$^{1, 4}$, and Ivan Mili\'c\,$^{3}$}

\def\Address{$^{1}$Laboratory for Atmospheric and Space Physics, University of Colorado, Boulder, CO 80303, USA \\
$^{2}$Department of Astrophysical and Planetary Sciences, University of Colorado, Boulder, CO 80309, USA \\
$^{3}$Institut für Sonnenphysik, Freiburg im Breisgau, Germany \\
$^{4}$Center for Astronomy, Space Science and Astrophysics, Independent University Bangladesh, Dhaka, Bangladesh}

\def\corrAuthor{Corinne Morrell}
\def\corrEmail{corinne.morrell@colorado.edu}

\begin{document}
\onecolumn
\firstpage{1}

\title[Local source wavefront detection]{Assessment of DKIST/VTF Capabilities for the Detection of Local Acoustic Source Wavefronts} 

\author[\firstAuthorLast ]{\Authors} 
\address{} 
\correspondance{} 

\extraAuth{}
\maketitle

\begin{abstract}

\section{}
Recent studies have demonstrated that temporal filtering can successfully identify local-acoustic-source wavefronts in radiative magnetohydrodynamic simulations of the solar photosphere. Extending this capability to observations promises new insight into the stochastic excitation of solar p-modes, the source depth distribution below the photosphere, and the dominant physical processes underlying acoustic wave excitation. Such measurements would also enable improved characterization of the complex wavefield in the lower chromosphere and open the possibility of ultra-local helioseismic diagnostics. In this work, we assess an observational strategy for the detection of local acoustic wavefronts on the Sun using the Visible Tunable Filter (VTF) instrument on the National Science Foundation's Daniel K. Inouye Solar Telescope (DKIST). Because wavefront identification requires high spatial and temporal resolution and is limited by the small amplitudes of the wave perturbations, we focus on identifying specific wavelength combinations within spectral lines that maximize the sensitivity to the wave signal. Under the cadence and spectral resolution constraints of DKIST/VTF observations and for the particular simulated wavefront we examine, this approach suggests two possible strategies: fast monochromatic imaging at 6302.425\,\AA, or ordered interleaved observations in the blue wing of either the Fe I 6302.5\,\AA\  or Fe I 5250.6\,\AA\ line (between 6302.419\,\AA\ and 6302.465\,\AA, or between 5250.579\,\AA\ and 5250.607\,\AA\ respectively).

\tiny
 \keyFont{ \section{Keywords:} Solar oscillations, Wave propagation, Forward modeling, Imaging spectroscopy, High-cadence observations, Helioseismology} 
\end{abstract}

\section{Introduction}\label{sec:intro}

There are two broad classes of stellar oscillations: high amplitude pulsations and low amplitude oscillations.
Self-excitation by overstability is thought to account for the former in stars such as classical Cephieds, RR Lyrae stars, $\delta$ Scuti variables, and pulsating white dwarfs, among others~\citep{2002ASPC..259...21C, 2010aste.book.....A, 2015EPJWC.10101002D, refId0}. By contrast, low-amplitude oscillations are typically associated with pulsationally stable stars, such as the Sun. Solar acoustic oscillations ($p$-modes) have photospheric radial velocity amplitudes of order $0.1$–$1~\mathrm{m\,s^{-1}}$, depending on the observable and the degree of spatial averaging, with associated intensity fluctuations of a few parts per million~\citep[e.g.,][]{2002RvMP...74.1073C}. No known mechanism can limit overstable pulsational modes to such low amplitudes, and continuous stochastic excitation is thought to sustain these oscillations~\citep{1977ApJ...212..243G}.

 Differing theoretical treatments emphasize the importance of different aspects of the underlying compressible turbulence to stochastic excitation. Convective motions may transfer energy into the acoustic wave spectrum via Reynolds stresses \citep[e.g.,][]{1952RSPSA.211..564L, 1954RSPSA.222....1L, 1967SoPh....2..385S, 1977ApJ...212..243G, 1990ApJ...363..694G, 1992MNRAS.255..639B, 2019ApJ...872...34K}, entropy fluctuations, \citep[e.g.,][]{1991LNP...388..195S, 1994ApJ...424..466G, 1997ASSL..225..135R, 1998IAUS..185..199N, 1999ApJ...524..462R, 2001A&A...370..136S, 2001A&A...370..147S}, or some combination of the two. More than one mechanism may be important, with the phase relationships between them possibly reflecting the evolution of solar granulation as the acoustic source~\citep{1999ApJ...524..462R}. In particular, both modeling~\citep[e.g.,][]{2000ApJ...541..468S,2020SoPh..295...26L} and observational~\citep{1995ApJ...444L.119R, 1998ApJ...495L..27G, 2000ApJ...535.1000S, 2010ApJ...723L.175R, 2010ApJ...723L.134B, 2013JPhCS.440a2044L} evidence suggests that acoustic excitation on the Sun occurs preferentially in granular downflow lanes and is associated with strong impulsive downflow events.  These may occur as a result of new downflow formation during granule fragmentation~\citep{1993ApJ...408L..53R, 1995ApJ...443..863R} or by the rapid enhancement of existing downflows.
 
 Acoustic emission in both cases includes dynamically-induced entropy fluctuations and Reynolds stresses ~\citep{1991LNP...388..195S, 1994ApJ...424..466G, 1997ASSL..225..135R, 1998IAUS..185..199N, 1999ApJ...524..462R, 2001A&A...370..136S, 2001A&A...370..147S}. In the case of granule fragmentation, the role of entropy fluctuations is enhanced by strong localized cooling~\citep{1997ASSL..225..135R, 1999ApJ...524..462R}. Although uncertainties remain due to the possible presence of a correlated noise component in the spectra \citep[][but see \citealt{2020A&A...635A..81P}]{1997MNRAS.292L..33R, 1998ApJ...495L.115N}, these excitation mechanisms are broadly consistent with constraints inferred from helioseismic phase-difference spectra. In particular, the measured velocity–intensity phase relationships agree with the spectral signatures expected from rapidly forming downflow plumes \citep{2000ApJ...535..464S, 2001ApJ...561..444S}. Other possible mechanisms for global $p$-mode excitation include solar flares, which can trigger transient acoustic waves \citep{1998Natur.393..317K, 2003SoPh..218..151A, 2005ApJ...630.1168D, 2020ApJ...895L..19M, 2020ApJ...901L...9L}, with coupling to solar $p$-modes somewhat uncertain but plausible~\citep{2014SoPh..289.1457L, 2016MNRAS.456.1826H}. 

Thus, multiple acoustic excitation mechanisms may be active on the Sun and in other stars.  Observational access to individual source events
(beyond those caused by flares)
would allow for greater constraints to be placed on the turbulent source depth distributions, the temporal phasing between the thermodynamic and dynamic fluctuations associated with the sources, and the coupling efficiencies to the global eigenmodes.  Linking the observed wave field to specific generation sites would clarify distinct excitation pathways and their relative contributions, enable improved characterization of the complex wavefield propagating upward from the photosphere into the lower chromosphere, and open the possibility of ultra-local helioseismic diagnostics based on the local wave field emitted by the discrete sources.
While current local helioseismic techniques leverage primarily the local wavefield of the global p-modes~\citep[e.g.,][]{2005LRSP....2....6G}, the focus here is on exploiting the near-field of individual sources to deduce subsurface properties on granular and subgranular scales.  As an example, this would allow for the diagnosis of magnetic field production during granule fragmentation~\citep{2018ApJ...859..161R}.

\subsection{The aim of this paper} \label{sec:intro_inversion}

In previous work, we have developed a diagnostic temporal differencing method designed to isolate small-amplitude acoustic perturbations generated by localized excitation events~\citep{2021ApJ...915...36B, 2023ApJ...955...31B}. We subsequently applied this method to numerical simulations to aid in the interpretation of early DKIST observations of transient wave phenomena~\citep{2024ApJ...971L...1B}. Based originally on a neural network designed to identify the local wave response to a discrete source, the method employs a third-order forward temporal finite-difference operator that, due to the difference in time scales, both suppresses the background granulation contributions and amplifies the signal of the rapid wavefront passage across the spatial grid.

The method was formulated and validated using a three-dimensional radiative magnetohydrodynamical simulation of quiet-Sun granulation~\citep{2014ApJ...789..132R}, operating directly on the temperature, pressure, and velocity fields computed by the simulation. The reliability of the differencing method, at face value, thus depends critically on having access to one or more of these physical quantities with high fidelity. Direct application to observations would require inferring the three-dimensional thermodynamic structure of the solar atmosphere from spectroscopic or spectropolarimetric measurements through inversion codes~\citep[e.g.,][]{1996SoPh..164..169D, 2016LRSP...13....4D}. In practice, however, this direct approach is limited by photon noise, finite spectral, spatial, and temporal resolution, radiative transfer degeneracies, and the regularization inherent to inversion procedures, all of which preferentially suppress small-scale structure and rapidly evolving perturbations, including the wavefronts of interest here. Instead, we explore an alternative strategy that does not aim to reconstruct the full atmospheric state but focuses on identifying the wavelengths in commonly observed photospheric lines at which the acoustic wavefront signal is most effectively captured.

Using the same numerical simulation in which we have already identified propagating wavefronts, we perform spectral synthesis to determine the wavelengths at which wavefront-induced intensity fluctuations are maximized relative to background variability. Spectral response functions provide a natural quantitative framework for this analysis, as they describe how perturbations at different heights in the atmosphere map into wavelength-dependent intensity variations. Although acoustic waves are traditionally characterized through pressure perturbations in simulations and Doppler velocity measurements in observations, the detectability of these weak transient signals in spectra is primarily governed by the temperature sensitivity of a spectral line. Accordingly, our analysis focuses on the temperature response functions, which directly quantify how depth-dependent thermal perturbations associated with the acoustic wavefronts translate into observable intensity variations. By evaluating the response of the synthesized spectra to the height-dependent temperature fluctuations induced by a wavefront passage, we construct a detectability metric that quantifies the relative contributions of the wave signal and of the background and identifies the spectral regions most favorable for wavefront detection using the successive temporal differencing technique.

Detection and characterization of local source wavefronts places stringent and often competing requirements on spatial resolution, temporal cadence, and spectral fidelity. Acoustic wavefronts associated with localized excitation events evolve on time scales of seconds, have spatial scales of tens of kilometers, and are most visible in the simulations as they propagate across a granule.  Thus, observing them with spectroscopic raster-scanning is very likely not possible. Any sequential sampling in physical space mixes temporal evolution with spatial structure, effectively smearing or suppressing the signatures of interest. Integral-field spectroscopy or fast Fabry–Pérot-based spectral line scanning (at a limited number of wavelengths) is therefore preferred, as these preserve the instantaneous spatial coherence of the wavefront while retaining spectral sensitivity. In the latter case, finite time is required to switch between wavelengths, and the image time series at different wavelengths, constructed by sequentially stepping through a limited set of line positions (and thus atmospheric heights), are not strictly co-temporal. We anticipate (and discuss further in Sections~\ref{sec:wavefronts} and~\ref{sec:wavesignatures}) that since the wavefront perturbations in which we are interested propagate upward in the atmosphere, this temporal offset can be accounted for, and used beneficially, in observation and data analysis strategies (Section~\ref{sec:obs}). Because of its combination of ultra-high spatial and spectral resolution, rapid cadence, and full-field spatial sampling, we suggest that the National Science Foundation's Daniel K. Inouye Solar Telescope \citep[DKIST,][]{2020SoPh..295..172R} Visible Tunable Filter \citep[VTF,][]{2016SPIE.9908E..4NS, Kentischer_2025_natvtf} is well positioned to meet the observational requirements of this problem (Section~\ref{sec:obs}). 

\section{Wavefront detection}
\label{sec:wavefronts}

\begin{figure}[t!]
\vskip 0.4cm
\centerline{\includegraphics[scale=0.45]{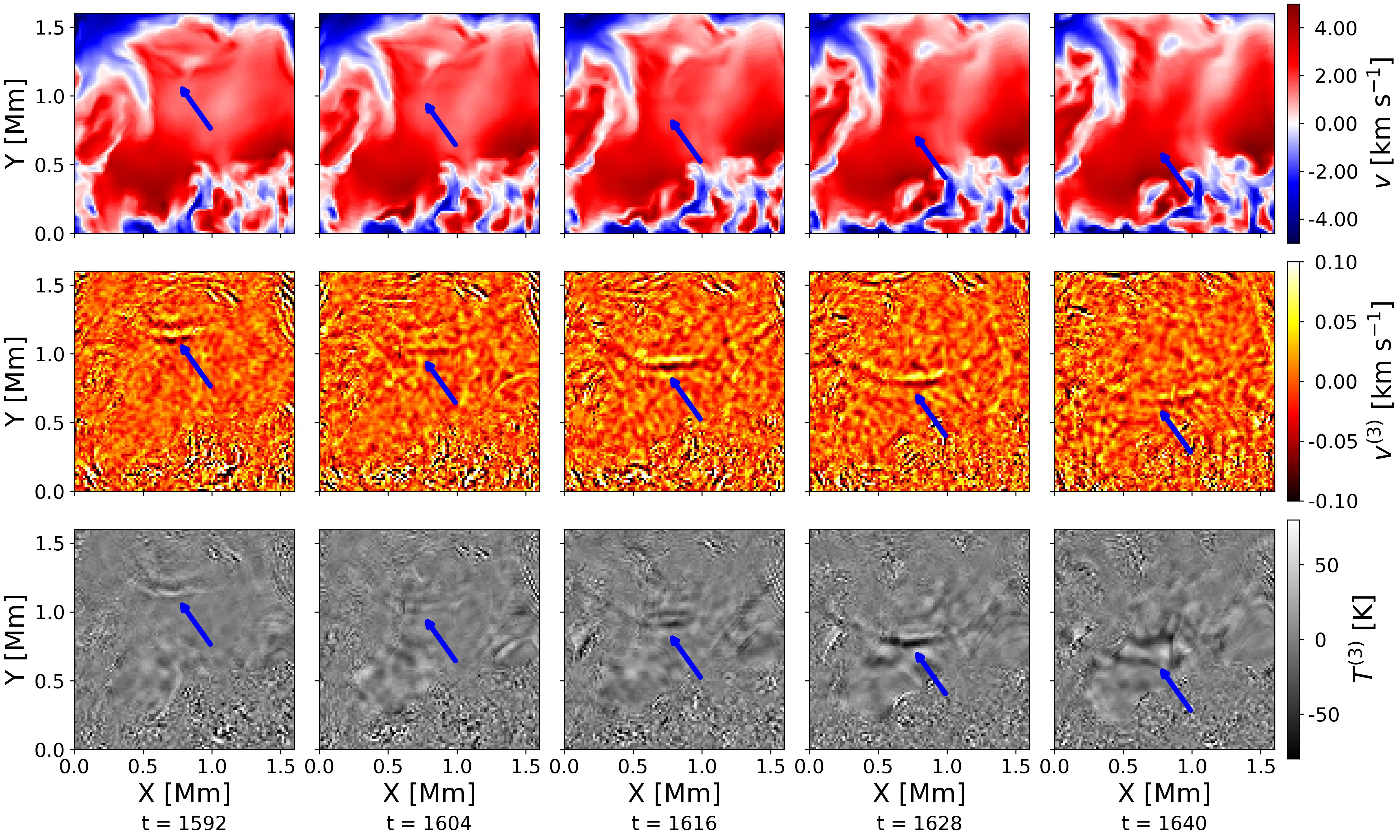}}
\caption{Propagation of a localized acoustic wavefront through the photosphere in a radiative MHD simulation. Panels show snapshots of the \textbf{(A) line-of-sight velocity} ({\it top}), its \textbf{(B) third-order temporal difference} ({\it middle}), and the corresponding \textbf{(C) third-order temporal difference of temperature} ({\it bottom}), revealing a coherent wavefront expanding across a granulation cell. Temporal differencing suppresses background convective motions and isolates the expanding wavefront as it traverses a granule, demonstrating the basis of the detection technique used throughout this work. Plotting cadence (sub-sampled for display) is 12\,s, while the temporal difference is applied to the native simulation cadence of 2\,s.  The original MURaM data cubes $(N_x, N_y, N_z) = (384, 384, 256)$ are cropped to a $100 \times 100 \times 100$ pixel sub-region ($1.6 \times 1.6$ Mm horizontally). The sub-region shown in this image is centered at pixel $(x_c, y_c) = (48, 119)$, where $(x_c, y_c)$ denote global indices in the full MURaM cube.}
\label{fig:wavefront}
\end{figure}

Our analysis is based on a MURaM quiet-Sun, mixed-polarity small-scale dynamo radiative magnetohydrodynamic (MHD) simulation of the upper solar convection zone, photosphere, and lower chromosphere~\citep[$O16b$ in][]{2014ApJ...789..132R}. The computational domain of the simulation spans 1 hour of solar time over a $6 \times 6 \times 3$\,Mm field of view with a grid spacing of 16\,km and a 2\,s cadence. This spatial and temporal resolution is sufficient to resolve propagating acoustic wavefronts produced by local sources. In the simulated photosphere, weak, small-scale magnetic fields are concentrated in intergranular lanes with nearby opposite polarities, with no coherent magnetic structure on the scale of plage. In the upper layers of the domain, the field becomes increasingly horizontally inclined, and the targeted photospheric wavefront signatures are primarily affected by variability in the convective background not the magnetic field.

In previous work, we identified 25 wavefronts of sufficient diagnostic quality (i.e., semicircular outwardly propagating perturbations visible over at least 6 or 7 timesteps) in the simulation~\citep{2023ApJ...955...31B}. These typically originate from shallow sources located some tens of kilometers beneath the photosphere. As the wavefronts pass through the photosphere, their horizontal phase speeds are initially very high and decrease with time as they expand, to asymptotically approach the sound speed at the source depth~\citep{2023ApJ...955...31B}. Once identified in velocity difference images, the propagating wavefronts are often just barely visible without temporal differencing in animated time series of the raw Doppler or temperature perturbation maps.

\begin{figure}[t!]
\centerline{\includegraphics[scale=0.45]{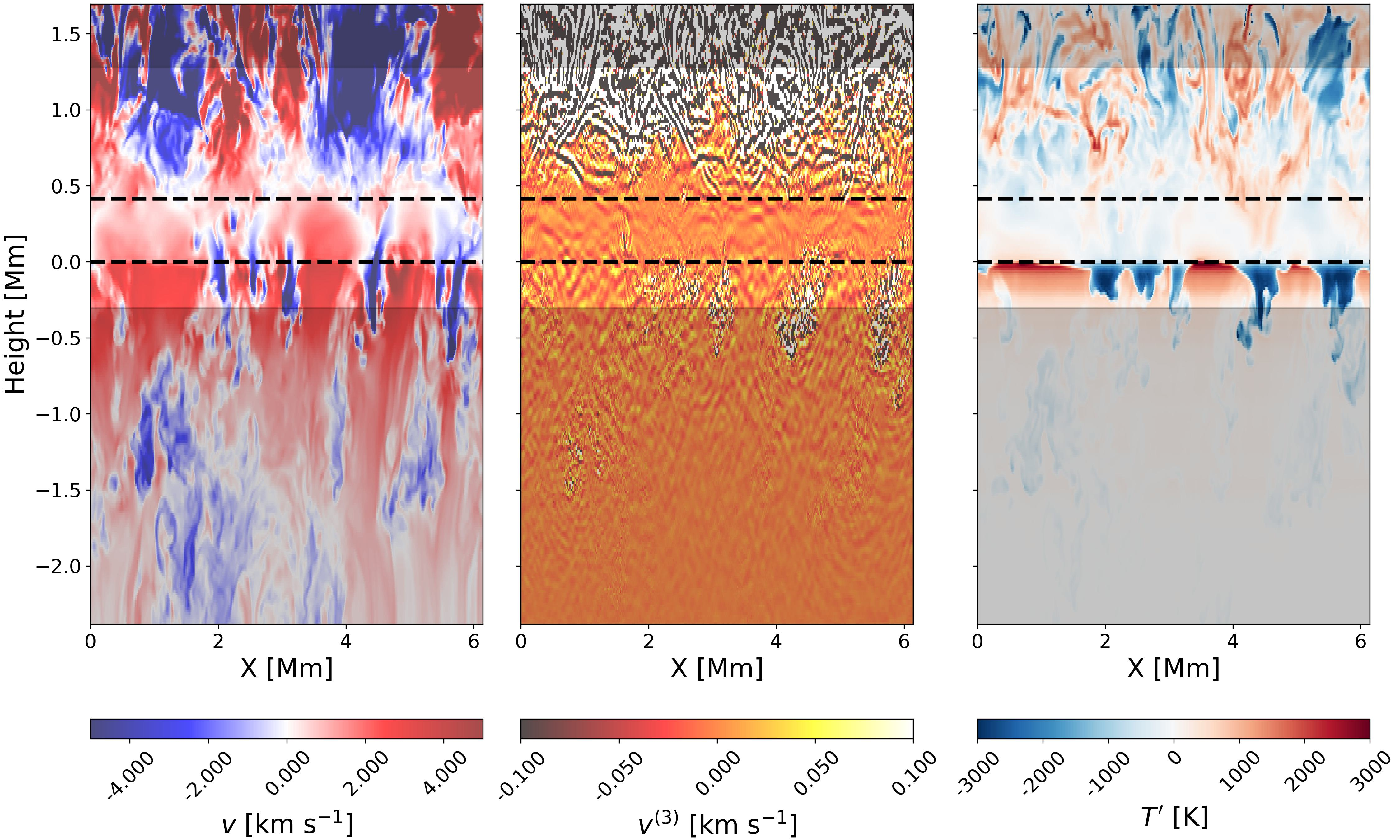}}
\caption{Height-dependent velocity and temperature structure of the simulated atmosphere at a single time step. The panel shows an instantaneous vertical slice of \textbf{(A) line-of-sight velocity $v$} ({\it left}), \textbf{(B) temporal velocity difference $v^{(3)}$} ({\it middle}), and \textbf{(C) temperature fluctuations $T^\prime = T - T_0$}, where $T_0$ is the horizontal mean temperature at each height ({\it right}), over the full vertical extent ($3\,$Mm) of the MURaM solution. The {\it dashed black} fiducial lines indicate the most wave-sensitive height range, $z\sim0.0 - 0.42$\,Mm (discussed in Section~\ref{sec:wavefronts}), since the background atmosphere in this region is quietest. The {\it shaded gray} regions indicate the layers excluded during spectral synthesis (Section~\ref{sec:spectra}).}
\label{fig:verticaldomain}
\end{figure}

The wavefront we analyze in detail in this paper occurs between 1584--1682\,s in the simulation and is the same event analyzed in \cite{2023ApJ...955...31B}. Figure~\ref{fig:wavefront} shows the wavefront propagating through the photosphere. Its visibility is particularly evident after applying the temporal differencing method to the Doppler velocity ($v^{(3)}$, {\it second} row) and the temperature field ($T^{(3)}$, {\it bottom} row). However, the visibility of the wavefront is not uniform with height in the simulated atmosphere because the amplitude of the wavefront relative to the background convective motions (which we consider here as noise) varies with height.

Figure~\ref{fig:verticaldomain} illustrates the background motions in which the wavefront is embedded, showing the vertical velocity (A), the vertical velocity temporal difference (B), and the temperature fluctuations about the horizontal mean (C), as functions of height at a single instant of time.  The region of convective overshoot, here taken as $z\sim0.0$ to $0.42$\,Mm above the photosphere (delimited with horizontal {\it black dashed} fiducial lines), has the lowest amplitude background noise. Convective amplitudes increase below the photosphere ($z\lesssim  0.0$), and the domain is filled with large amplitude acoustic gravity waves above $z\sim0.42$.

Time series observations of high-resolution photospheric spectra in quiet-Sun regions \citep[e.g.,][]{2005A&A...441.1157P} show similar overshoot signatures extending through the lower photosphere, with the granule-lane temperature contrast becoming small around $\sim$0.15-0.17\,Mm and vertical velocity structures remaining detectable up to heights of order $\sim$0.37\,Mm.  Therefore, assuming the simulation in this respect accurately represents the Sun, observations aiming to uncover impulsively generated wavefronts have the highest likelihood of success if they are focused on the solar convective overshoot region. We note that, since the simulation domain is quite shallow, the upper boundary condition may play some role in establishing the chromospheric wave field. However, a wave field of similar amplitude is also found in deeper simulations~\citep[those of][]{rempel2025} which have an upper boundary that extends into the corona. The origin, persistence, and importance of this wavefield to the chromospheric structure and dynamics are being pursued but are not the focus of this paper.

\section{Spectral signatures}
\label{sec:spectra}

The first step in quantifying the spectral sensitivity to local wavefront perturbations is the synthesis of the emergent Stokes~$I$ intensity from the MURaM simulation. Given that the heights of interest (see Figure~\ref{fig:verticaldomain}) lie in the region between the base of the photosphere and the temperature minimum (the region from about 0-0.5~Mm above mean $\log\tau=0$), we closely examine three photospheric lines, synthesizing Fe~I~5250.2\,\AA, Fe~I~5250.6\,\AA, and Fe~I~6302.5\,\AA, over wavelength ranges 5249.90--5250.45\,\AA, 5250.45--5251.00\,\AA, and 6302.20--6302.90\,\AA\ with 0.005 \AA\ sampling. All three lines are commonly chosen for their magnetic sensitivity (not central to our study) and have been observed with imaging spectropolarimeters such as CRISP \citep{Scharmer_2008_CRISP} and TuMag \citep{2025SoPh..300..148D}.
In our synthesis, we consider only vertical rays and restrict the vertical extent of the atmosphere to the \textit{unshaded} region in Figure~\ref{fig:verticaldomain} ($-\!0.30\le z\le\,1.28$\,Mm).  The latter restriction reduces the computational cost of the transfer calculation while still capturing the full range of heights that contribute to absorption in the lines of interest. Using the temperature, pressure, and line-of-sight velocity profiles in a small region of interest (see Section \ref{sec:perturbations} below),
we synthesize emergent spectra for each atmospheric column.  We employ the \texttt{SNAPI} code \citep{2018A&A...617A..24M}, solving the polarized radiative transfer equation assuming Local Thermodynamic Equilibrium (LTE). Thus, we obtain $I(\lambda,t)$ for each atmospheric column at each horizontal location in the region of interest.

\subsection{Response}
\label{sec:rf}

The sensitivity of the emergent intensity to atmospheric perturbations is quantified through the spectral response function. Of particular interest for wavefront detection is the temperature response function (${\rm RF}_T$),
which measures the linear changes in the emergent intensity at a given wavelength produced by localized perturbations in $T$ as a function of perturbation height $z$.  
\texttt{SNAPI} computes the derivatives of the specific intensity with respect to changes in $T$ by evaluating variations in the formal solution that result from small perturbations to the atmospheric column at each height. The result is a two–dimensional kernel, 
\begin{equation}
\mathrm{RF}_{T}(\lambda,z)
    = \frac{\partial I(\lambda)}{\partial T(z)}\ ,
\label{eq:rf}
\end{equation}
for each atmospheric column at each time step in the simulation.

The response functions for the wavelength ranges we consider peak between $z \approx 0\,$Mm and $ 0.5\,$Mm. Because the acoustic wavefronts in the simulation are most apparent as small amplitude fluctuations in the overshoot region, which lies within this height range, these response functions provide a natural bridge between the physical perturbations induced by the wavefront passage and their observational signatures.

\subsection{Perturbation definition and wavefront mask}
\label{sec:perturbations}

\begin{figure}[t!]
\vskip 0.4cm
\centerline{\includegraphics[scale=0.45]{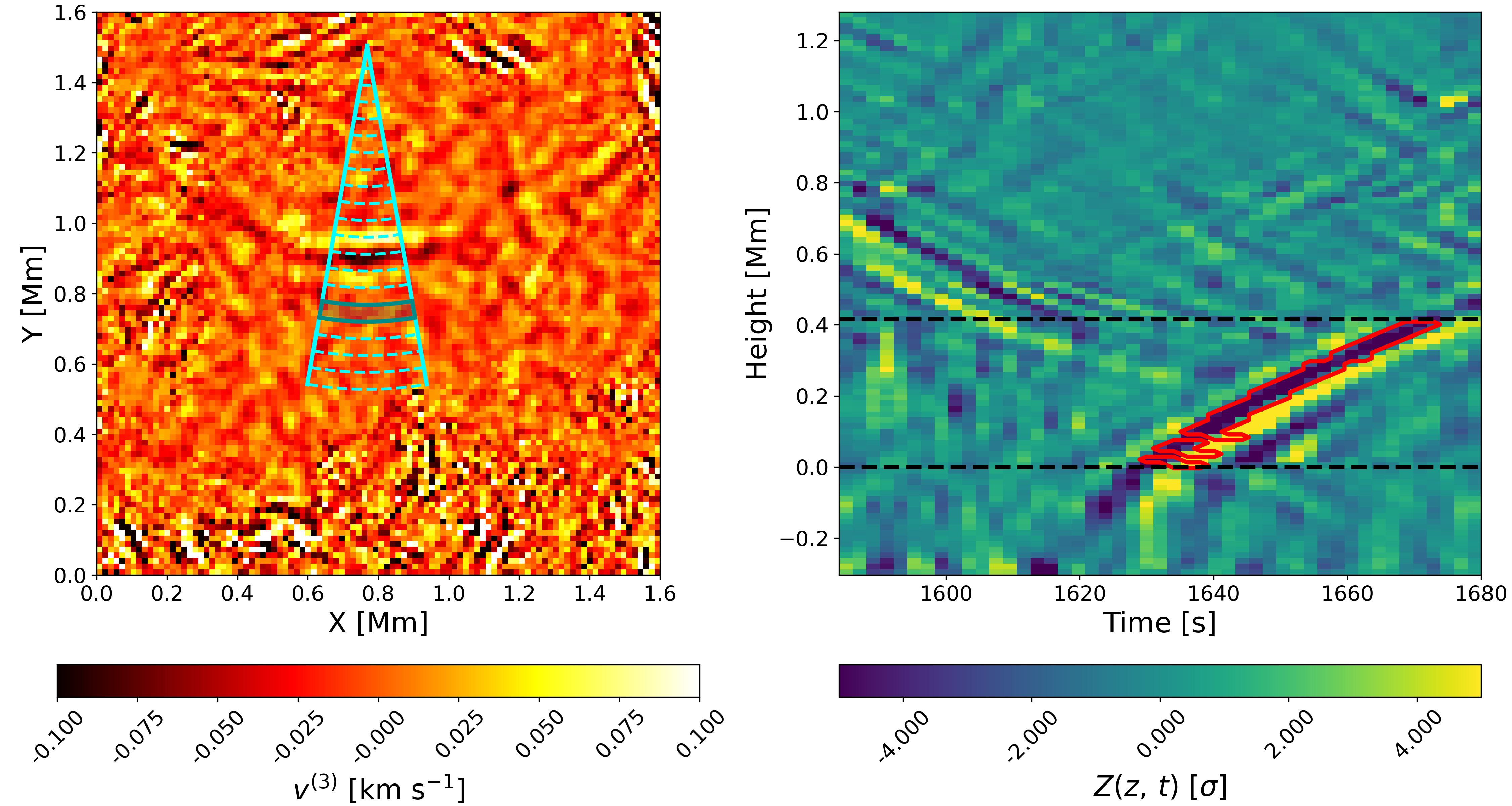}}
\caption{Construction of azimuthally averaged acoustic mask from the MURaM simulation.  \textbf{(A) Temporal velocity difference $v^{(3)}$} snapshot ({\it left}) showing the acoustic wavefront in the MURaM photosphere with concentric annuli around fitted center  (\textit{cyan}). Annulus selected for analysis indicated with dark, bolded boundaries. \textbf{(B) Standard score $Z(z,\,t)$ map} for the region of interest ({\it right}), revealing a clear, height-dependent acoustic propagation pattern. The wavefront becomes clearly detectable in the selected annulus at 1632\,s and remains visible for 50\,s. The simulation interval 1584--1682\,s is chosen for spectral synthesis to include a pre-event interval of comparable duration with no visible wavefront signal. \textit{Dashed black} lines indicate the most wave-sensitive height range while the \textit{solid red} line outlines the region isolated by the mask described in Section \ref{sec:perturbations}.}
\label{fig:region-of-interest}
\end{figure}

With the goal of determining where in the spectrum the acoustic wavefront will produce the largest observable signature, we first locate the wavefront in the simulation by manually tracing its position in the $v^{(3)}$ (velocity difference) images at the photosphere ($z=0$\,Mm). We measure its angular extent and locate the center of its expansion by fitting semicircular arcs to its shape. Using these measurements, we define the region of wavefront propagation  (the wedge in Figure~\ref{fig:region-of-interest}A) and divide it into concentric annular segments with fixed radial widths of 3 pixels (48 km), which is the approximate extent of the wavefront in the direction of propagation in the simulations.

For each annulus, we compute the $v^{(3)}$ standard score ($Z$-score) at each height and time and define the wave as present when $|Z|>3$. The annulus with the strongest integrated $Z$-score over height and time is selected here for further analysis and is highlighted with bold boundaries in Figure~\ref{fig:region-of-interest}A. A time-height map of the $Z$-score for this annulus is shown in Figure~\ref{fig:region-of-interest}B, where the upward-propagating wavefront is clearly visible in the convective overshoot region, the region between $0.0\le z\le 0.42$ and bounded by horizontal fiducial lines in the figure.  As an aside, the figure also shows evidence of a weaker possible downward propagating feature that may be relevant to observations of downward propagating waves in the solar atmosphere~\citep[e.g.,][and references therein]{2022A&A...665L...2G,2024ApJ...966..200C} and perhaps related to shock dissipation~\citep{2024ApJ...971L...1B}, but it has not yet been analyzed in detail and is not the subject of this paper.

\begin{figure}[t!]
\vskip 0.4cm
\centerline{\includegraphics[scale=0.75]{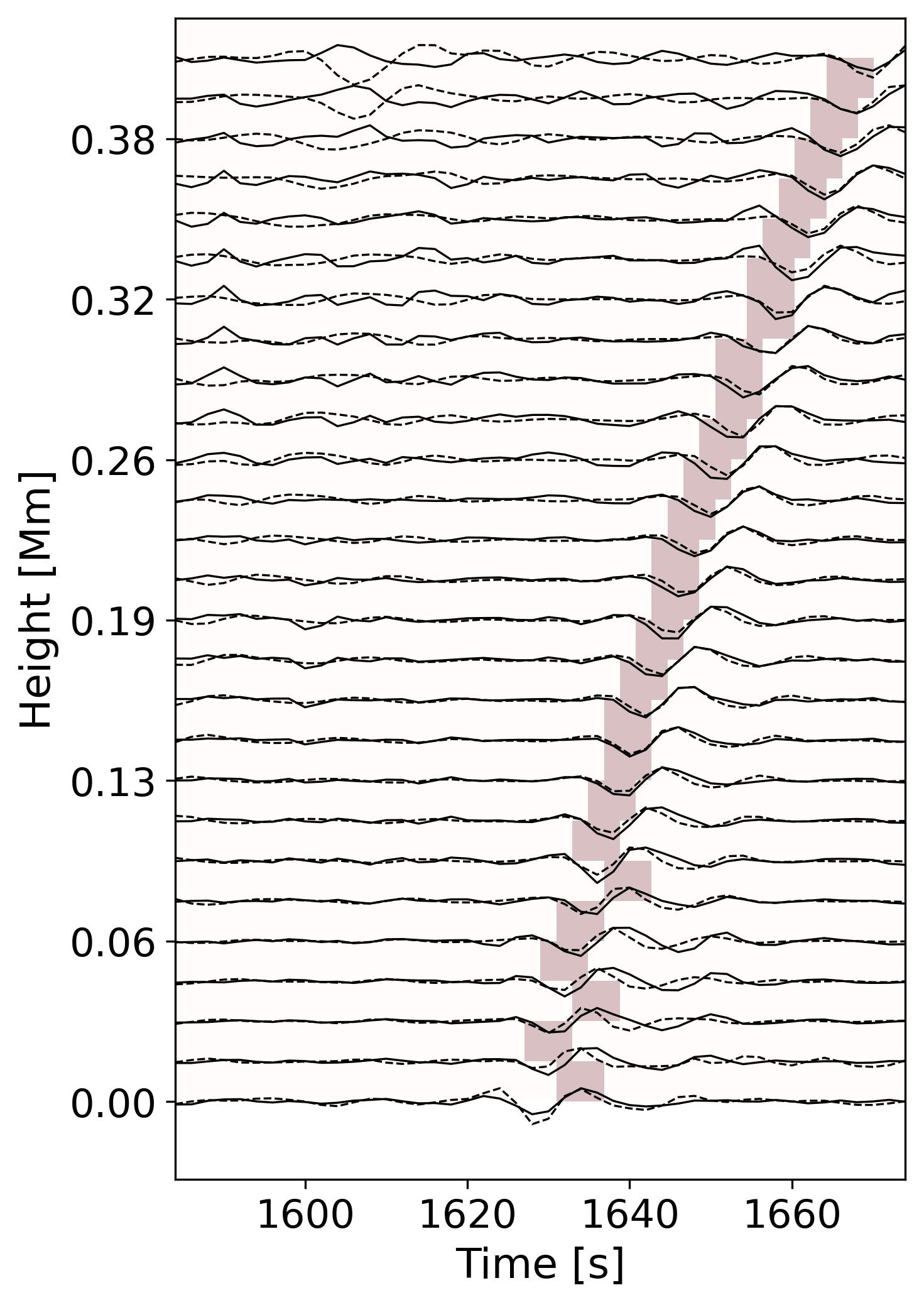}}
\caption{One-dimensional upward wave propagation constructed from azimuthally averaged perturbations in the annulus of interest. The temporally differenced velocity $v^{(3)}$ (\textit{solid black} line) and differenced temperature perturbation $\delta T^{(3)}$ (\textit{dashed black} line) are shown as functions of time and vertically offset by height, revealing the coherent upward progression of the acoustic wavefront through the atmosphere. The \textit{shaded red} area indicates the wave mask $M_{\mathrm{wave}}$ of Section \ref{sec:perturbations} used to isolate the wave-dominated portion of the signal.}
\label{fig:stacked_signals}
\end{figure}

Within the selected annulus, we compute the azimuthally averaged temperature $\bar{T}(z,t)$ at each height $z$ and time $t$ and define the perturbation relative to a wave-free reference as

\begin{equation} \label{eq:Tpert}
\delta T(z,t) = \bar{T}(z,t) - \bar{T}_{\mathrm{bg}}(z),
\end{equation}
where $\bar{T}_{\mathrm{bg}}$ is the annular mean temperature at each height averaged over a wave-free time interval (simulation time period $1504-1560\,$s, ending approximately $30\,$s before the wavefront is visible in any annulus in the photosphere). Analogous calculations are performed for Doppler velocity and pressure.

To isolate the temperature perturbation associated with the wavefront, we define a mask $M_{\mathrm{wave}}(z,t)$ by selecting the time steps surrounding the $Z$-score peak ($\pm 2\,$s) at each height. This approximates the time it takes for the acoustic perturbation to cross 3 grid cells ($48\,$km), which is about the thickness of the wavefront in the direction of propagation. The mask is shown as a {\it red} outline in Figure~\ref{fig:region-of-interest}B and is superimposed as a shaded region in Figure~\ref{fig:stacked_signals}. Although the masked region looks like a trough in the temporally differenced variables, it physically corresponds to a region of positive (upward) velocity and positive temperature perturbation.

Using this mask, we thus define the characteristic temperature perturbation at the wave location as a function of height and time as

\begin{equation}
    \delta T_{\mathrm{wave}}(z,t) = \delta T(z,t) \cdot M_{\mathrm{wave}}(z,t)\, .
\label{eq:dT}
\end{equation}
We note that the temperature perturbation $\delta T_{\mathrm{wave}}(z,\,t)$ is defined relative to the mean quiet background state, and thus does not explicitly separate the wavefront contributions from others such as granulation. As the background thermodynamic structure evolves in time, $\delta T_{\mathrm{wave}}(z,\,t)$ does not represent a purely wave-isolated perturbation but the total deviation from the mean quiet reference atmosphere when the wave is present.

Subsequent analysis focuses on the temperature perturbations caused by the wavefront passage because they contribute most strongly to changes in the spectrum. However, we define the location of the wavefront using $v^{(3)}$ because it is most easily identified in the simulations using this measure, and accurately locating the wavefront is critical to our assessment of its observability. For this excitation event, the wave-induced temperature and Doppler velocity perturbations are in phase at the wavefront location (Figure~\ref{fig:stacked_signals}). Thus, the location of the wavefront based on $v^{(3)}$ is a consistent proxy for the location of the wave-induced temperature perturbations. The temperature-velocity phase relation found here is somewhat surprising in light of previous studies of acoustic waves on the Sun, which report a phase difference between velocity and temperature or intensity~\citep[e.g.,][]{1990IAUS..138..251M, 1997A&A...324..704S, rast_asymmetry_1998, 1999ApJ...516..939S, 2000ApJ...535..464S, 2001ApJ...561..444S, 2023ApJ...952...58V}. Those previous results pertain to continuously driven modal oscillations and atmospheric wave fields for which the phase relation can be different from that of the local response to a discrete impulsive source.

\subsection{Sensitivity Measure}
\label{sec:snr}

\begin{figure}
    \centering
    \includegraphics[scale=0.5]{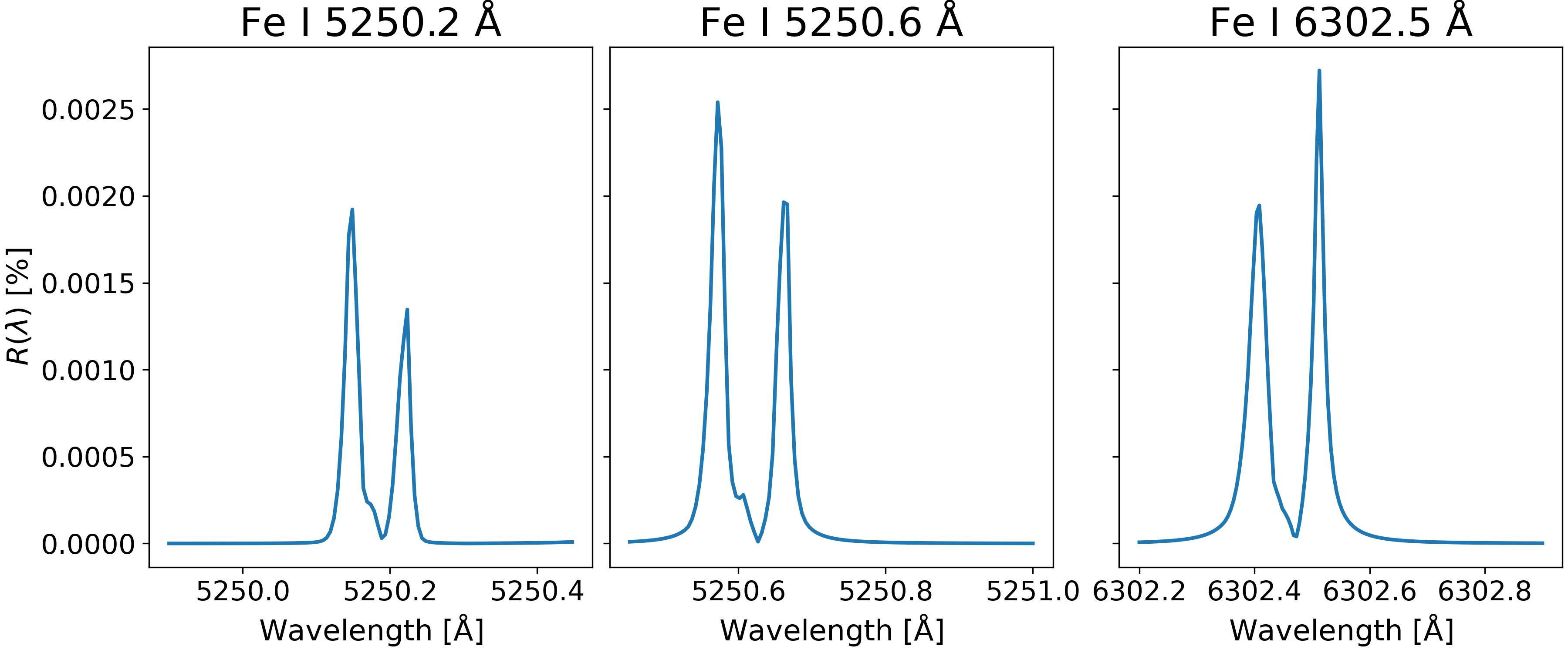}
    \caption{Ratios of the amplitudes of the maximum intensity perturbations due to velocity and temperature fluctuations, i.e., $R(\lambda) = \mathrm{max}_t[\Delta I_v(\lambda, t)]/\mathrm{max}_t[\Delta I_T(\lambda, t)]$, for the specific acoustic event considered here.}
    \label{fig:velocity_contribution}
\end{figure}

For small amplitude perturbations, the change in emergent intensity at each height in the simulation at a given wavelength $\lambda$ is approximately
\begin{equation}
\Delta I(\lambda,t)
    = \int \left[\mathrm{RF}_{T}(\lambda,z)\,\delta T(z,t) + \mathrm{RF}_{v}(\lambda,z)\,\delta v(z,t)
    + \mathrm{RF}_{p}(\lambda,z)\,\delta p(z,t)\right]\,dz\ .
\label{eq:linpert}
\end{equation}
We exclude contributions from velocity and pressure fluctuations in our analysis because the temperature contribution to $\Delta I$ dominates for the spectral lines and wavefront event we consider here (see Figure \ref{fig:velocity_contribution}).

With Equation~\ref{eq:linpert}, we can define a wavelength-dependent sensitivity measure $\mathrm{SNR}(\lambda, t)$ to assess at which wavelengths the wavefront signal most exceeds that due to typical background fluctuations. 
The wave-related contributions to the intensity changes $\Delta I(\lambda,t)$ are

\begin{equation}
    \Delta I_{\mathrm{wave}}(\lambda,t) = \int \mathrm{RF}_{T}(\lambda,z)\,\delta T_{\mathrm{wave}}(z,t)\,dz\ ,
\end{equation}
where the integral is effectively over the limited range of heights over which the wave is found due to the mask that has been applied to obtain $\delta T_{\mathrm{wave}}$ (Equation~\ref{eq:dT}).
The background noise contribution to $\Delta I(\lambda,t)$ can be expressed as

\begin{equation}
    \Delta I_{\mathrm{std}}(\lambda,t) = \int \mathrm{RF}_{T}(\lambda,z)\,\sigma_T(z,t)\,dz,
\end{equation}
where $\sigma_T(z,t)$ is the standard deviation of the temperature fluctuations at each height and the integral extends over all heights. We measure the standard deviation of the temperature fluctuations over the annulus of interest (Figure~\ref{fig:region-of-interest}A) and all times over which the spectral synthesis was performed (1584--1682 s).  

\begin{figure}[t!]
\vskip 0.4cm
\centerline{\includegraphics[scale=0.45]{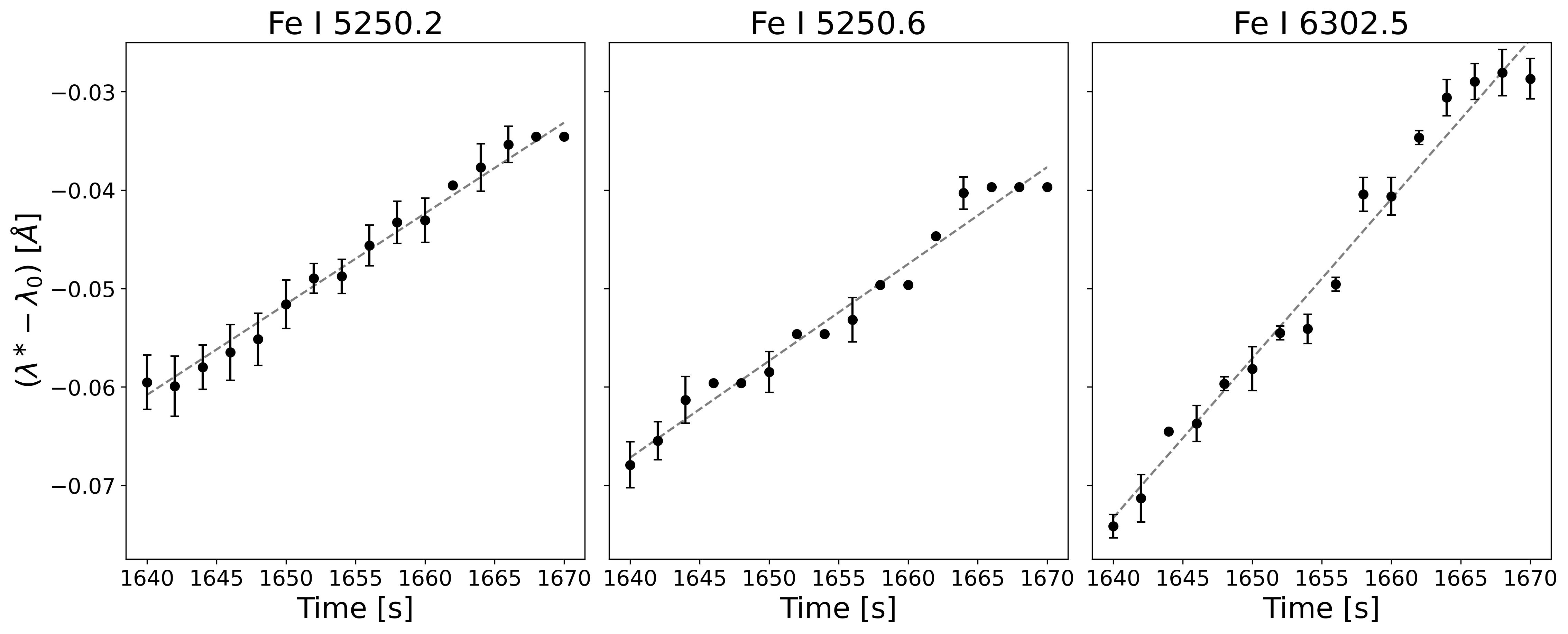}}
\caption{Time-dependent drift of the wavelength of maximum temperature sensitivity, $\lambda^{\ast}$, averaged over the annulus of interest. The evolution of $\lambda^{\ast}$ toward the rest-frame line center ($\lambda_0$) traces the upward propagation of the acoustic wavefront and the corresponding shift in the atmospheric layer contributing most strongly to the intensity response. Error bars indicate the standard deviation of $\lambda^{\ast}$ over pixels.}
\label{fig:lambda_star_v_zstar}
\end{figure}

The wavefront sensitivity measure is then

\begin{equation}
\mathcal{S}(\lambda,t)=   
\frac{\Delta I_{\mathrm{wave}}(\lambda,t)}{\Delta I_{\mathrm{std}}(\lambda,t)}\ ,
\label{eq:snr}
\end{equation}

and the wavelength of maximum sensitivity is

\begin{equation}
\lambda^\ast(t) = \underset{\lambda}{\arg\max}\;\mathcal{S}(\lambda,t)\ .
\end{equation}
As with the response function (Equation~\ref{eq:rf}), the wavefront sensitivity measure $\mathcal{S}(\lambda,t)$, and thus $\lambda^\ast(t)$, is computed independently for each pixel in the annulus of interest. The average $\lambda^\ast(t)$ over that annulus is plotted as function of time in Figure~\ref{fig:lambda_star_v_zstar}, and represents the wavelength at which the characteristic signal from the wavefront has the largest amplitude compared to variations at that wavelength due to fluctuations at all heights.

We define a second quantity of interest, $\lambda^{**}$, as the spatial average (over the annulus of interest) of the wavelength at which the sensitivity measure is highest at any time. For each pixel within the annulus of interest, we find the wavelength at which the wave sensitivity is maximum when considering all wavelengths and all times within the wave mask. We report a single $\lambda^{\ast\ast}$ for each spectral line by spatially averaging this quantity, yielding the wavelength at which the wavefront is, on average, most detectable.

\begin{figure}[t!]
\vskip 0.4cm
\centerline{\includegraphics[scale=0.45]{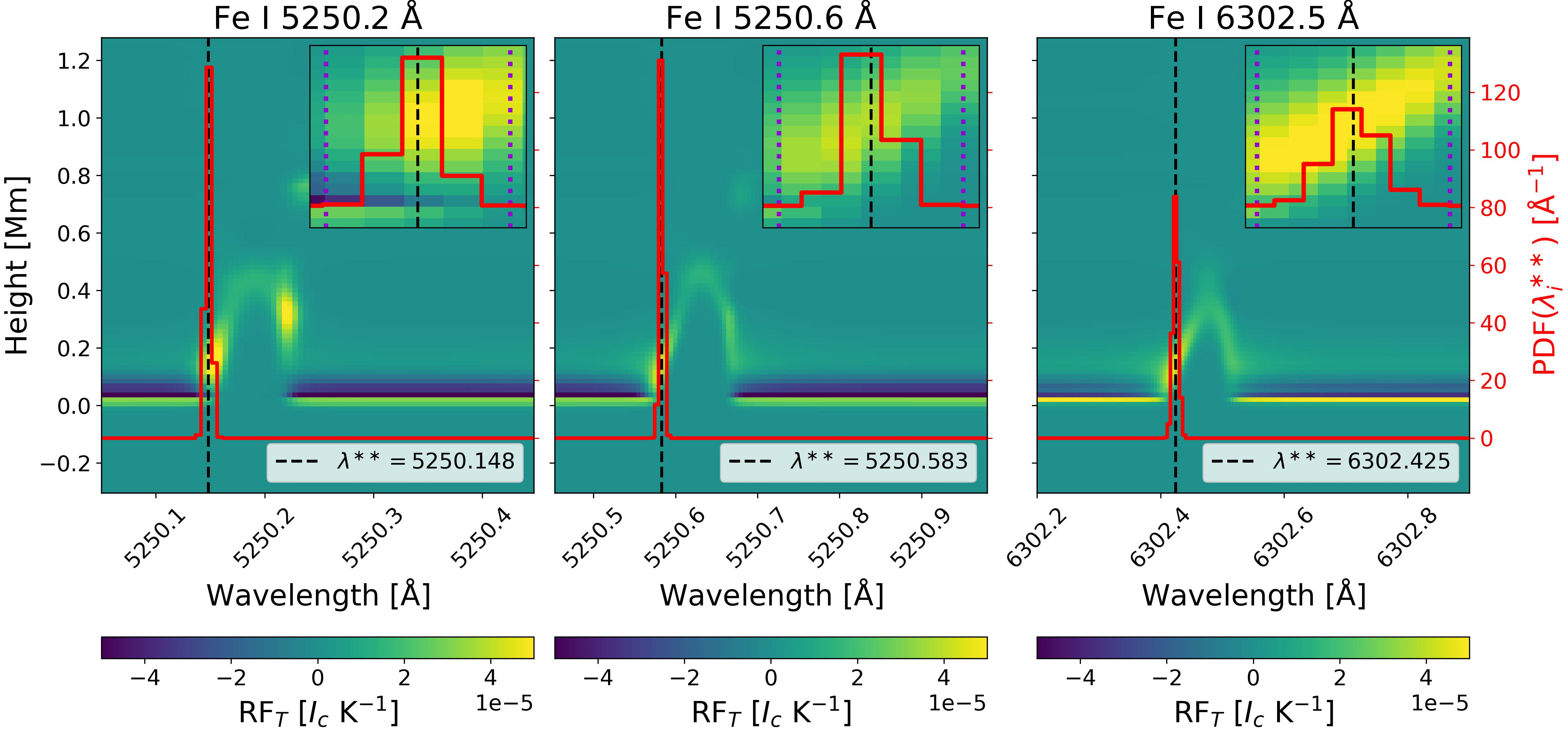}}
\caption{The average, over the annulus of interest, of the temperature response function $\mathrm{RF}_T$ (normalized by continuum intensity $I_c$) for each atmosphere making up the annulus of interest at $t = 1640\,$s, across the Fe I 5250.2, 5250.6, and 6302.5 spectral lines ({\it background}) The probability density function of $\lambda^{\ast\ast}$ for each pixel in the annulus of interest is over-plotted in {\it red}, with $\lambda^{\ast\ast}$ indicated with a {\it black} vertical line. The insets zoom into the $\lambda^{\ast\ast}$ region and indicate the VTF spectral element extent with {\it dotted} vertical lines.}
\label{fig:lambda_star_PDF}
\end{figure}

The distribution of wavelength values contributing to this mean (the individual pixel values) is shown in Figure~\ref{fig:lambda_star_PDF}. The values found for $\lambda^{**}$ and their standard deviations are
$\lambda^{**}=5250.148\,$\AA\ , $5250.583\,$\AA, $6302.425\,$\AA\ , and 
$\sigma_{\lambda^{**}}=0.0031$\,\AA, $0.0028$\,\AA, $0.0046$\,\AA\, for the Fe I 5250.2, Fe I 5250.6, and Fe I 6302.5 lines, respectively.  The time at which the sensitivity measure $\mathcal{S}(\lambda,t)$ reaches its maximum (i.e., when we measure the largest change in the emergent intensity as the wavefront passes) is $t=1642$\,s, corresponding to a wavefront height between $z=0.128$ and $0.176$ Mm above the photosphere (above the mean $\tau=1$ height in the simulation). Because all three lines considered belong to the same species and exhibit comparable line strengths that lead to relatively similar sensitivities, perturbations to their intensities evolve similarly as the wavefront propagates through the atmosphere. As a result, the sensitivity measure peaks at the same time for all three lines.

\subsection{Wavefront signatures}
\label{sec:wavesignatures}

For all lines examined, maximum spectral sensitivity occurs on the blue (shorter wavelength) side of the line, due to the systematic Doppler distortion caused by the granular upflow underlying the region of wavefront propagation. As shown in Figure~\ref{fig:lambda_star_PDF}, this distortion reduces the slope of the response function, resulting in a relatively narrow range of heights which contribute to the emergent intensity. This, in turn, increases the sensitivity measure $\mathcal{S}(\lambda, t)$ given in Equation~\ref{eq:snr}. Thus, line profiles with less steep wings are generally preferred because they localize the response function in height and reduce contamination from noise-dominated regions of the atmosphere.

As shown in Figure~\ref{fig:lambda_star_v_zstar}, maximum spectral sensitivity shifts towards the line core as the wavefront propagates upward in the atmosphere. The rate of this systematic shift does not simply reflect the wavefront propagation speed.  Instead, $\lambda^\ast(t)$ is determined by the atmosphere through which the wave is traveling (defining the response function), the vertical propagation speed of the wavefront, and the spectral slope of the line wing.  It can be used in the development of a robust observing strategy (Section~\ref{sec:obs}), either by fitting $\lambda^\ast(t)$ for each spectral line using the simulation or, more satisfactorily, by determining from first principles how $\lambda^\ast$ evolves given the spectral line and the atmosphere of interest.  The latter is the focus of future efforts.  Here we measure the slopes and find them to be $9.223\times10^{-4}$, $9.845\times10^{-4}$, and $1.621\times10^{-3}\,$\AA\ s$^{-1}$\, for Fe I 5250.2, Fe I 5250.6, and Fe I 6302.5 respectively.

\section{Observational Strategies -- Implications for DKIST/VTF}
\label{sec:obs}

The two measures, $\lambda^{*}(t)$ and $\lambda^{**}$, motivate different observing strategies.
The first is an estimate for the best wavelength sequence to employ to follow the wavefront propagation, while the second is an estimate of the best overall choice of wavelength at which to perform monochromatic imaging at high cadence. Observing at $\lambda^{**}$ is roughly equivalent to observing the wavefronts as they propagate horizontally at the height of maximum sensitivity.  Observing sequentially over the wavelengths $\lambda^{*}(t)$, on the other hand, allows collection of a data cube that can be post-processed to find wavefront signatures as they propagate upward with height. For a more robust measurement, the individual $\lambda^{*}$ maps in multi-wavelength observations could be shifted in time, stacked, and summed to obtain the mean signal over all heights. In this operation, the wave signal adds coherently, while the variation in granulation with height contributes to some cancellation of the background and consequent enhancement of the propagating wavefront signal-to-noise.

The VTF instrument provides full-field imaging spectroscopy across $5200–8700$\, \AA, with a $60^{\prime \prime} \times 60^{\prime \prime}$ field of view sampled at $0.014^{\prime\prime}$ per pixel, corresponding to about $10.5$ km on the solar surface. This sampling is comparable to the 16 km horizontal grid spacing of the MURaM simulation analyzed for this paper, and allows access to the spatial scales over which acoustic wavefronts remain coherent and may be recoverable by temporal differencing~\citep{2023ApJ...955...31B}.
The spectral resolving power of VTF is $R\sim10^5$ (e.g., $\sim$0.06 \AA\ resolution with $\sim$0.03 \AA\ sampling near 6000 \AA) over a free spectral range of $\sim$10 \AA. In detail,  the VTF spectral sampling width is about 0.023\,\AA\ at 5250\,\AA\ and 0.033\,\AA\ at 6302\,\AA~\citep{2016AN....337.1064T,2016SPIE.9908E..4NS}.  These band passes are over-plotted in the insets of Figure~\ref{fig:lambda_star_PDF}. 

As discussed above, the distributions of $\lambda^{**}$, as determined from the individual atmospheres within the annulus of interest, have widths of $\sigma_{\lambda^{**}}=0.0031$\,\AA, $\sigma_{\lambda^{**}}=0.0028$\,\AA, and $\sigma_{\lambda^{**}}=0.0046$\,\AA\, for Fe I 5250.2, Fe I 5250.6, and Fe I 6302.5, respectively.  Thus, the range of $\lambda^{**}$ values found for each of these lines lies well within one spectral sampling element of the VTF.  VTF observations centered on $\lambda^{**}$ or $\lambda^{*}$ will capture a relatively broad range of wavelengths and consequently sample a range of atmospheric heights. It is clear from Figure~\ref{fig:lambda_star_PDF} that, in the case of Fe I 5250.2 for example, centering the VTF passband on $\lambda^{\ast \ast}$ will include significant contributions from the continuum.  Those contributions are dominated by granulation noise that will likely overpower the wave signal in monochromatic observations.  More generally, the range of atmospheric heights encompassed by the VTF bandpass may smear the wave signal. However, the blue wing of Fe I 5250.6, and even more so that of Fe I 6302.5, are shallower and thus less contaminated by contributions away from the wavelength of maximum sensitivity (see insets of Figure~\ref{fig:lambda_star_PDF}).  Here we focus our observational plans on the Fe I 6302.5 line, though careful convolution with both the
VTF spectral bandpass and the DKIST point spread function (PSF) is required to fully assess the potential of each line for local wavefront discovery.

\subsection{Monochromatic Observations}
\label{sec:mono}

To comply with the requirements of the temporal difference method, the cadence of the monochromatic observations should be at or below the critical sampling rate $\Delta t = \Delta x /v_{\mathrm{ph}}$, where $\Delta x$ is the spatial sampling width and $v_\mathrm{ph}$ is the phase speed of the wavefront. The phase speed of the wavefront at $z=0.152\,$Mm, the midpoint of the Fe I 6302.5 $\lambda^{**}$ height, is $~10.5$ km/s.  This speed is estimated from the frame-to-frame radial advance of the wavefront measured with the azimuthally averaged acoustic mask (Figure~\ref{fig:region-of-interest}) and the mean value over the interval in which the front traverses the granule is quoted.  Critical sampling of the wavefront thus occurs at a cadence of 1 s. The practical viability of this cadence is governed by the signal-to-noise requirements based on the intensity fluctuations caused by acoustic perturbations. 

For intensity-based diagnostics, the relevant signal-to-noise ratio is defined as
\begin{equation}
\mathrm{SNR} \equiv \frac{I}{\sigma_I},
\end{equation}
where $I$ is the measured intensity and $\sigma_I$ is the corresponding uncertainty. Detecting a fractional intensity perturbation $\Delta I / I$ with $N_\sigma$ significance requires
\begin{equation}
\mathrm{SNR} \gtrsim \frac{N_\sigma}{\Delta I / I}.
\end{equation}
In practice, this implies the following order-of-magnitude requirements: a $1\%$ intensity fluctuation requires $\mathrm{SNR} \sim 100$ for marginal ($1\sigma$) detectability and $\sim 300$ for robust ($3\sigma$) detection; a $0.5\%$ fluctuation requires $\mathrm{SNR} \sim 200$--$600$; and a $0.1\%$ fluctuation requires $\mathrm{SNR} \sim 300$ for marginal detection and 1000 for robust detection.

Quantitative guidance on the achievable cadence vs. signal-to-noise trade space is assessed by the VTF Instrument Performance Calculator~\citep[IPC,][]{ipcweb}. For monochromatic (single-wavelength) intensity imaging, the IPC indicates that an SNR of 235 can be achieved at a cadence of 0.04\,s. Relaxing the cadence to 1.04\,s, however, increases the SNR to 1200. While our analysis is for a single representative wavefront, the phase speed of the wavefront depends on the depth of the excitation source, with deeper sources producing higher phase speeds. Because the source depth in the solar atmosphere is not known a priori, and constraining it is one of the primary goals of the observations we propose, the cadence must accommodate a plausible range of propagation speeds. For wavefront propagation speeds between 7\,km\,s$^{-1}$ (the lower bound of the photospheric sound speed) and 11\,km\,s$^{-1}$ (the upper range observed in the simulations analyzed here), sampling intervals between 0.95 and 1.5\,s satisfy the minimum sampling requirement. Therefore, we suggest monochromatic sampling at $\lambda^{**}$ with a cadence of 0.5 s, which over-samples the expected wavefront propagation while maintaining a high SNR of approximately 817. Post-processing can further increase the effective SNR through temporal averaging (e.g., a running average).

\subsection{Multi-wavelength Observing Strategy}
\label{sec:multi}

For multi-wavelength observations, the goal is to scan the range of wavelengths $\lambda^\ast(t)$ that track the upward propagation of the wavefront through the atmosphere. The wavelengths chosen must span a significant portion of that range of $\lambda^{*}(t)$ and follow the slope of $\lambda^{*}$ with time (Figure~\ref{fig:lambda_star_v_zstar}). The scan across all wavelengths must be completed quickly enough to allow repeat measurements at any single wavelength with a cadence that allows temporal differencing of the time series.  To accomplish this without knowledge of the  wavefront location, a data volume in $(x,y,\lambda, t)$ must be assembled from which the trajectories can be extracted.

The wavelength of maximum sensitivity $\lambda^{*}(t)$ in the blue shoulder of Fe I 6302.5 (Figure~\ref{fig:lambda_star_v_zstar}) extends from 6302.419\,\AA\  to 6302.465\,\AA\ as the wavefront propagates upward in the atmosphere over a time period of 30\,s, with a slope of $1.621\times10^{-03}\,$\AA\,s$^{-1}$.    
The VTF IPC imposes a minimum wavelength step between successive observations of 0.0315\,\AA. Observations sampling $\lambda^{*}(t)$ would ideally be spaced closer in wavelength but, under this restriction, the strategy is to interleave observations over three wavelengths, $\lambda^{**}$ and $\lambda^{**}\pm 0.0315$, i.e., 6302.3935, 6302.425, and 6302.4565\,\AA.  This can be done with a cadence of 0.48\,s achieving a signal to noise at each wavelength of 408. Since the endpoints of this wavelength triplet lie slightly outside of the $\lambda^{*}(t)$ range, and thus have lower wavefront sensitivity, an alternative two wavelength option is warranted, observing the lower end of the $\lambda^{*}(t)$ range at 6302.42\,\AA\ and 6302.45\,\AA.  This two-wavelength observation can be done with a cadence of 0.48\,s with a signal to noise of 527. The wavefront transit time between the atmospheric heights sampled by these two wavelengths is about 18.5\,s.

An observation period of 60\,s with excellent seeing would allow approximately two full wavefront transit times to be sampled under the first multi-wavelength scheme and approximately three under the second.  Longer excellent seeing time series would further improve the sampling statistics.  In the MURaM photosphere, we observe about 25 propagating wavefront events per hour over the $36\,{\rm Mm}^2$ field.  Thus, we expect to observe about 40 events per minute over the full $60^{\prime \prime} \times 60^{\prime \prime}$ VTF field of view (FOV). This estimate is an upper limit that depends on the detection criteria and assumes maximum possible spatial resolution over the whole FOV, which is not guaranteed. However, even with an order of magnitude fewer events, we are still poised for multiple detections per minute.

As discussed above, post-observational analysis of the multi-wavelength data cube is critical and should focus on using the height-time dependencies of wavefront trajectories to increase detection likelihood. This can be done by looking directly for propagating signals in the four-dimensional space-wavelength-time data cube or by shifting the data accumulated at different wavelengths and times so that they can be added together to reinforce the signal.  Since the data is to be acquired at a cadence of $0.48\,$s and the propagation time across the sampled wavelengths takes about 15 to 30\,s for the two observation scenerios, trajectories with different upward signal propagation speeds should be accessible and distinguishable in the data cube.

\section{Conclusions}

In this work, we have developed a physically motivated framework for detecting locally excited wavefronts in the solar photosphere by connecting the source wavefronts to their wavelength-dependent observational signatures in commonly observed Fe I spectral lines. This connection was established by determining the wavelengths of maximum sensitivity to the wavefront perturbation as it propagates upward through the overshoot region of a simulated solar atmosphere. Based on these synthetic results, we defined two observation strategies for the detection of these transient events.

A central conclusion of the work is that the measurement techniques we suggest are well fit to DKIST/VTF capabilities, though a smaller minimum wavelength separation between successive observations would be very beneficial. While the detection of expanding, spatially coherent wavefronts is  incompatible with raster-based spectroscopy, which entangles spatial structure with temporal evolution and thus confuses signals from fast transient events, the VTF’s Fabry–Pérot–based, full-field imaging spectroscopy preserves their coherence.  Moreover, the temporal offset between time series of interleaved wavelengths can be leveraged to enhance the propagating signal. VTF performance calculations suggest that, under the operational constraints of the instrument, sparse wavelength sampling concentrated in the most sensitive line-wing regions can satisfy the temporal requirements imposed by signal propagation while still retaining sufficient spectral information to track the upward evolution implied by the drift of the optimal sensitivity wavelength. 

At the same time, the approach carries inherent limitations that arise directly from its design. First, the usability of the wavefront identification technique depends sensitively on the stability and alignment of successive image frames.  Wavefront identification is based on detecting coherent, rapidly evolving perturbations against the more slowly varying background, such that small residual differences between consecutive frames can produce a difference signal. Although post-image reconstruction \citep[e.g.,][]{vanNOort_2005_momfbd} can help, it cannot remove image misalignment and instability entirely. Therefore, these effects impose practical limits on the minimum detectable wave amplitude.  A comprehensive test of the whole data acquisition process, including simulating time-dependent atmospheric seeing and an end-to-end test of the image reconstruction for VTF-like data, would help assess the impact of image stability on wavefront detection, but such a test goes beyond the scope of this paper.

Second, the critical requirement for sub-second to second temporal cadence to robustly separate the wavefront signal from the noise by temporal differencing constrains the number of wavelength positions that can be sampled within a single scan. Consequently, the method trades spectral coverage for temporal sampling. Combined with practical restrictions on the minimum wavelength separation between successive observations, this limits the ability to perform detailed atmospheric inversions or to disentangle overlapping contributions from different heights. 

Third, spectral resolution and stability play a central role in determining the vertical localization of the measurement. Our current method relies on isolating wave-induced perturbations over a narrow and stable wavelength range, as encoded in the wavelength-dependent response functions of the spectral line. If the instrument resolution or stability worsens, contributions from a wider range of atmospheric heights are mixed within a single wavelength sample, reducing the contrast of the wavefront signal relative to background fluctuations. In this regard, spectral lines with soft (not steep) wings more reliably localize measurements in height. Our preliminary tests show that the Na~I~D lines at 5888~\AA\ and 5896~\AA\ exhibit such features, thanks to their broad wings as compared to the Fe spectral lines considered in this paper. Technical observations in one of the Na lines were already conducted as a part of the VTF's first light \citep{Schubert_2025_aas}. Given their non-LTE nature, the Na~I~D lines require more detailed modeling, which is currently in progress.

Despite these limitations, our assessment indicates that the DKIST 4\,m aperture, in combination with the VTF Fabry–Pérot imager, constitutes one of the few observational platforms capable of simultaneously satisfying the spatial, spectral, and temporal resolution constraints required by the methods suggested here. Consequently, local acoustic wavefront detection represents both a compelling scientific objective and a stringent benchmark for DKIST's ability to probe weak, rapidly evolving photospheric dynamics at their intrinsic spatial and temporal scales. Successful application of the framework developed would transform a long-standing problem into an observationally accessible diagnostic.

\section*{Conflict of Interest Statement}
The authors declare that the research was conducted in the absence of any commercial or financial relationships that could be construed as a potential conflict of interest.

\section*{Author Contributions}
CM: Conceptualization, Formal analysis, Methodology, Visualization, Writing -- original draft; MR: Conceptualization, Interpretation, Methodology, Supervision, Writing -- original draft; SB: Conceptualization -- original project conception, Methodology, Writing -- original draft; IM: Software, Writing -- review \& editing.

\section*{Funding}
This work was partially supported by the National Science Foundation (NSF) Award No. 2206589. CM acknowledges support from the University of Colorado Hale Fellowship program and the NSF Graduate Research Fellowship Program (GRFP), Grant No. DGE-2040434.

\section*{Acknowledgments}
Sincere thanks to M. Rempel for his continued and ongoing scientific generosity. The simulation material is based upon work supported by the NSF National Center for Atmospheric Research, which is a major facility sponsored by the U.S. National Science Foundation under Cooperative Agreement No. 1852977.  The authors thank the anonymous referees for their careful reading and valuable input.

\section*{Materials and Methods}
The analysis presented in this study is based on a three-dimensional radiative magnetohydrodynamic simulation of quiet-Sun convection produced with the MURaM code, spanning the upper convection zone, photosphere, and lower chromosphere. Localized acoustic wavefronts are identified using a temporal differencing technique applied to the simulated velocity and temperature fields, followed by azimuthal averaging to isolate coherent wave-induced perturbations from background convective variability. Synthetic spectral profiles are computed using the SNAPI radiative transfer framework, and temperature response functions are used to quantify the wavelength-dependent sensitivity of spectroscopic intensity to wave-driven thermal perturbations. All subsequent detectability metrics are derived from these response functions and annulus-averaged perturbations, enabling direct assessment of optimal spectral sampling strategies for DKIST/VTF observations.

\section*{Data Availability Statement}
The simulation data and analysis products supporting the findings of this study are available from the corresponding author upon request. The numerical simulation was produced using the MURaM code, and all post-processing, spectral synthesis, and analysis procedures were carried out using SNAPI code and custom python script described in the text. Derived data products and scripts necessary to reproduce the figures and results can be provided by the corresponding author.

\bibliographystyle{Frontiers-Harvard}
\bibliography{references}




\end{document}